\documentclass[prl,twocolumn,showpacs,floatfix]{revtex4}
\usepackage{graphicx}

\begin{document}

\title{Ferromagnetic EuS films: Magnetic stability, electronic structure
  and magnetic surface states}

\author{Wolf M\"{u}ller}
\homepage{http://tfk.physik.hu-berlin.de}
\author{Wolfgang Nolting}
\affiliation{Lehrstuhl Festk\"{o}rpertheorie,
  Institut f\"{u}r Physik, Humboldt-Universit\"{a}t zu
  Berlin, Newtonstra{\ss}e 15, 12489 Berlin, Germany}
\pacs{71.27.+a,73.20.At,75.50.Pp,75.70.Ak}
\begin{abstract}
We present the temperature and layer dependent electronic structure of a
20-layer EuS(100)-film using a combination of first-principles and model
calculation, the latter based on the ferromagnetic Kondo-lattice. The
calculated thickness-dependent Curie temperature agrees very well with
experimental data. The projected 5d-bandstructure is at finite temperatures
strongly influenced by damping effects due to spin exchange
processes. Spin-split unoccupied 5d-surfaces states are found with a
Stoner-like collapsing for increasing temperature towards the Curie
point and with an exponential decay of spectral weight with increasing
distance from the surface.
\end{abstract}
\maketitle

Europiumsulphide is a prototype ferromagnetic semiconductor, crystallizing
in the rocksalt structure with a lattice constant $a=5.95$ {\AA}. The
$\mathrm{Eu}^{2+}$ ions occupy lattice sites of an fcc structure. The just
half-filled 4f shell of the rare earth ion creates a strictly localized
magnetic moment of 
$7\mu_{B}$ according to the ground state configuration
$^{8}S_{7/2}$. The moments are exchange coupled resulting in a
ferromagnetic order for temperatures below the Curie temperature
$T_{\mathrm{C}}=16.57 \mathrm{K}$ 
\cite{Wachter79}. As to the purely magnetic properties bulk-EuS is
considered rather well understood being an almost ideal realization of
the Heisenberg model, where the exchange 
integrals can be restricted to nearest ($J_{1}/k_{B}=0.221 \mathrm{K}$)
and next  
nearest neighbors ($J_{2}/k_{B}=-0.100 \mathrm{K}$), only
\cite{BZDK80}. 

Recently, there has been an increasing
interest in thin EuS films, in particular what concerns the thickness
dependence of the ferromagnetic transition temperature
\cite{stachow-wojcik99:_ferrom_eus_pbs} which is frequently discussed in
terms of finite-size scaling (dimensionality crossover)
\cite{Cape76}. Investigations of this kind have already earlier been
done for films of the metallic counterpart Gadolinium \cite{FBS+93}. In
addition, Gd has provoked numerous research activities with respect to
its extraordinary magnetic surface properties as, e. g., a possibly
enhanced Curie temperature of the (0001) surface compared to that of
bulk Gd \cite{WAG+85}. 

While the Heisenberg model provides an
excellent description of the purely magnetic properties of local-moment
insulators such as EuS, EuO or metals like Gd it is of course 
inadequate for describing
electronic and
magnetooptic effects. So it cannot explain the striking temperature
dependence of the 5d conduction bands, first observed for the
ferromagnetic europium compounds as red
shift of the optical absorption edge for electronic 4f-5d transitions
upon cooling below $T_{\mathrm{C}}$
\cite{Wachter79}. The reason is an interband exchange coupling
of the excited 5d electron to the localized 4f moments that transfers the
temperature dependence of the ferromagnetic moment state to the
unoccupied conduction band states. Recently the same temperature effects
have been verified for film structures in the case of EuO by spin-resolved x-ray absorption spectroscopy
\cite{Steen02} and by quasiparticle band structure calculations
\cite{schiller01:_kondo,schiller01:_predic_euo}. 
In ref.\ \cite{schiller01:_predic_euo} the theoretical prediction of a
magnetic surface state has led to the speculation that the temperature
dependence below $T_{\mathrm{C}}$ may give rise to a surface insulator-halfmetal
transition accompanied by a huge magnetoresistance effect. 
It is one of the goals of our present study to teach whether this
spectacular proposal may hold for EuS too.

We present in this paper a study of the temperature and layer dependent
electronic structure of thin EuS-films by combining a many-body model
with a first-principles bandstructure calculation. The final
goal are statements about magnetic stability in terms of the Curie
temperature and the existence of surface states. 

In the following we assume EuS film structures with two surfaces
parallel to the fcc(100) crystal plane and consisting altogether of d equivalent
parallel layers. The lattice sites $\mathbf{R}_{i\alpha}$ within the
film are indicated by letters $ \alpha$, $\beta$, 
$\dots$
denoting the layer, and by letters i, j, $\dots$ numbering the
sites within a given layer, for which we assume translational symmetry
in two dimensions. Each lattice site is occupied by a magnetic moment,
represented by a spin operator $\mathbf{S}_{i\alpha}$, due to the
half-filled 4f shell of the $\mathrm{Eu}^{2+}$ ion. The exchange coupled moments
(spins) are certainly well described by a Heisenberg Hamiltonian:
\begin{equation}
  \label{eq:Heisenberg}\textstyle
  H_{4\mathrm{f}}=\sum_{ij\alpha\beta}J_{ij}^{\alpha\beta}
  \mathbf{S}_{i\alpha}\mathbf{S}_{j\beta}-D_{0}
  \sum_{i\alpha}\left(S_{i\alpha}^{z}\right)^{2}   
\end{equation}
As mentioned the exchange integrals $J_{ij}^{\alpha\beta}$ can be
restricted to nearest ($J_{1}$) and next nearest ($J_{2}$)
neighbors. The dipolar energy of EuS films is taken into account by a
single-ion anisotropy $D_{0}$. This helps to overcome the Mermin-Wagner
theorem \cite{MW66} which forbids a collective moment order in films of
finite thickness with isotropic Heisenberg exchange
. Since the exchange constants are derived from a
low-temperature spin wave analysis \cite{BZDK80}, we fix $D_{0}$,
somewhat arbitrarily, by the requirement that our theory yields the
experimental $T_{\mathrm{C}}$ for bulk EuS. We find
$D_{0}/k_{B}=0.375\mathrm{K}$ 
\cite{MuNo02}.

For the 5d conduction electrons we use the partial operator $H_{5d}$:
\begin{equation}
  \label{eq:5dpart}\textstyle
  H_{5\mathrm{d}}=\sum_{ij\alpha\beta}\sum_{mm'}
  T_{ij\alpha\beta}^{mm'}c_{i\alpha 
    m\sigma}^{\dagger}c_{j\beta m'\sigma}
\end{equation}
$c_{i\alpha
    m\sigma}^{\dagger}$ ($c_{i\alpha
    m\sigma}$) is the creation (annihilation) operator of an electron
  with spin $\sigma$ at site $\mathbf{R}_{i\alpha}$ in the orbital
  m. The $T_{ij\alpha\beta}^{mm'}$ describe the electron hopping from
  $\mathbf{R}_{j\beta}$ to  $\mathbf{R}_{i\alpha}$ with a possible
  orbital change ($m'\rightarrow m$). We require that these
  single-electron energies do not only account for the kinetic energy and
  the influence of the lattice potential, but also for all those
  interactions which are not explicitly covered by the model
  Hamiltonian. Therefore we take them from an LDA calculation. Details
  are given below.

Many important properties of local-moment materials such as EuS can be
traced back to the interaction between the 4f and 5d partial
systems. Starting from the very general on-site Coulomb interaction
between electrons of different orbitals it can be shown 
\cite{schiller01:_kondo,schiller01:_temper_euo} for the special case of
EuS (half-filled spin-polarized 4f shell, empty conduction bands) that
the interband interaction can be written as:
\begin{eqnarray}
  \label{eq:dfinteraction}
  \lefteqn{H_{\text{4f-5d}}=-\textstyle\frac{1}{2}J
    \sum_{im\alpha}\big(
    S_{i\alpha}^{z}(n_{im\alpha\uparrow}-n_{im\alpha\downarrow})}
   \hspace{6em}\\
   &&\textstyle
   +S_{i\alpha}^{+}c_{im\alpha\downarrow}^{\dagger}c_{im\alpha\uparrow}  
   +S_{i\alpha}^{-}c_{im\alpha\uparrow}^{\dagger}c_{im\alpha\downarrow}\big)    \nonumber  
\end{eqnarray}
J is the corresponding exchange coupling constant. Furthermore, we have
used the standard abbreviations:
$n_{im\alpha\sigma}=c_{im\alpha\sigma}^{\dagger}c_{im\alpha\sigma}$;
$S^{\pm}=S^{x}\pm \mathrm{i}S^{y}$. The first term in (\ref{eq:dfinteraction})
describes an Ising-like interaction between the localized 4f spin and
the spin of the itinerant 5d electron, while the two others provide spin
exchange processes between the two subsystems. Spin exchange may happen
in three different elementary processes: magnon emission by an itinerant
$\downarrow$ electron, magnon absorption by an $\uparrow$ electron and
formation of a quasiparticle (\textit{magnetic polaron}). The latter can
be understood as a propagating electron dressed by a virtual cloud of
repeatedly emitted and reabsorbed magnons corresponding to a
polarization of the localized-spin neighborhood.

We believe that the total model Hamiltonian
\begin{equation}
  \label{eq:totalHam}
  H=H_{4\mathrm{f}}+H_{5\mathrm{d}}+H_{\text{4f-5d}}
\end{equation}
incorporates the main physics of the ferromagnetic local-moment
insulator EuS. It can be considered as the multiband version of the
ferromagnetic Kondo-lattice model (KLM). To get a realistic description
of EuS we try to cover all those interactions, which do not explicitly
appear in our model Hamiltonian, by a proper renormalization of the
single-particle energies. For this reason we performed a band-structures
calculation using the tight-binding LMTO-atomic sphere
approximation program of Andersen
\cite{andersen75,andersen84}. Difficulties typical of LDA arise with the
localized character of the 4f levels. A "normal" LDA calculation
produces wrong 4f positions.
To
circumvent the problem we consider all the seven 4f electrons as core
electrons in one spin channel, 
since the 4f levels enter our study only as localized spins in the sense
of $H_{\mathrm{4f}}$ in (\ref{eq:Heisenberg}). We have to choose the proper
single-particle input in such a way, that a double counting of just the
decisive interband exchange (\ref{eq:dfinteraction}), explicitly by the
model Hamiltonian (\ref{eq:totalHam}) and implicitly by the LMTO input,
is avoided. The most direct solution of this problem would be to switch
off the interband exchange $H_{\text{4f-5d}}$ in the LDA code, what turns out
to be impossible. We can exploit the fact that the non-trivial
many-body problem of the KLM is exactly solvable for a ferromagnetically
saturated semiconductor, e. g. EuS at $T=0\mathrm{K}$ \cite{SMN96}. The
result is especially simple for the $\uparrow$ spectrum because the $\uparrow$
electron cannot exchange its spin with the fully aligned 4f spins. Only
the Ising term in $H_{\text{4f-5d}}$ takes care for a rigid shift of the total
$\uparrow$ spectrum by $\frac{1}{2} JS$. Thus we
can use without any 
manipulation the $\uparrow$ dispersions of the LMTO calculation, which
by definition works for $T=0\mathrm{K}$, for the EuS-hopping integrals in
(\ref{eq:5dpart}). There is no need to switch off $H_{\text{4f-5d}}$ because in
this special case it leads only to an unimportant rigid shift. It should
be stressed, however, that the $\downarrow$ spectrum, on the contrary,
is strongly influenced by the interband exchange, even at $T=0\mathrm{K}$ where
it is still exactly calculable \cite{SMN96} (see fig.~4).

\begin{figure}[htbp]
    \includegraphics[clip=,width=0.635\linewidth]{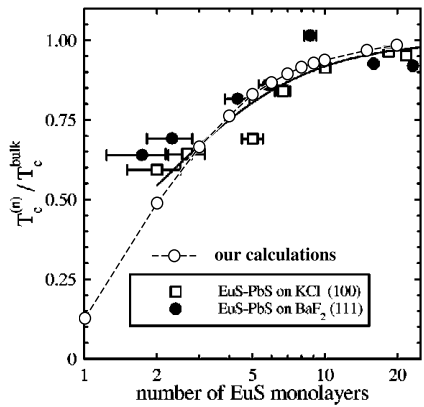}
    \caption{Curie temperature as function of film thickness (number of
      EuS monolayers). Our theoretical results: $\circ$ - - $\circ$. The
      experimental data are taken from
      ref.\ \cite{stachow-wojcik99:_ferrom_eus_pbs}. 
      $T_{\mathrm{C}}^{\mathrm{bulk}}=16.57\mathrm{K}$. 
    \label{fig:Curietemp}}
\end{figure}

Because of the absence of conduction electrons in the semiconductor EuS,
at least at low temperatures, the magnetic ordering of the localized 4f
moments will be unaffected by the band states, because an RKKY-type
contribution cannot occur. On the other hand the relevant superexchange
mechanism is hidden in the exchange integrals $J^{\alpha\beta}_{ij}$ in
(\ref{eq:Heisenberg}), which are taken from the experiment \cite{BZDK80}
The magnetic part of the
problem can therefore be solved separately via the extended Heisenberg
Hamiltonian $H_{\mathrm{4f}}$. We employ a generalized version of the
well-known Tyablikow approach%
, from which we know that it yields
convincing results 
in the low- as well as high-temperature region. Details of the method
are presented in \cite{schiller01:_temper_euo}. The result is
a distinctly layer- and temperature-dependent magnetization of the 4f moment
system. The Curie temperature of the EuS-film exhibits a strong
thickness-dependence, starting at about 2K for the monolayer and
approaching the 
bulk-value for more than 25 monolayers (Fig.~\ref{fig:Curietemp}). The
$T_{\mathrm{C}}$ of the 20 layer 
film, we are going to discuss below, amounts to 16.28K. We find a remarkable
agreement of our theoretical 
results with experimental data found by Stachow-Wojcik et
al.\ \cite{stachow-wojcik99:_ferrom_eus_pbs}. A similar $T_{\mathrm{C}}$
behavior, derived from susceptibility measurements,
is reported by Farle et al.\ \cite{FBS+93} for the metallic counterpart
Gd. The decrease of $T_{\mathrm{C}}$ 
with decreasing film thickness is understood as typical finite size
scaling accompanied by a dimensionality cross-over.

To obtain the temperature-dependent electronic structure of a EuS(100)
film the $T=0$-$\uparrow$ dispersions from our tight binding-LMTO band calculation are
needed as single-particle input for the model evaluation. To account for
the film geometry, we define a supercell consisting of n consecutive
EuS(100) layers followed by m layers of empty spheres, i. e. periodically
stacked EuS n-layer films, isolated from each other by m layers of empty
spheres. It turned out that $m=5$ is large enough to guarantee truly
isolated EuS films. We do not consider surface relaxation and
reconstruction, which might be present in EuS films.

Similarly to the magnetic part the electronic part can be treated
separately because magnon energies are smaller by some orders of
magnitude compared to other energy terms as the exchange coupling
constant J or the conduction band width. So we can disregard the moment
operator $H_{\mathrm{4f}}$ when calculating the electronic
self energy. That does 
not at all mean that the magnetic 4f moments do not influence the
electronic quasiparticle spectrum. That is rather done by local-spin
correlations such as $\langle S^{z}\rangle$, $\langle S^{\pm}S^{\mp}\rangle$,
$\langle(S^{z})^{2}\rangle$,$\cdots$, which are to a great extent
responsible for the 
temperature-dependence of the electronic spectrum and have to be
derived from $H_{\mathrm{4f}}$. To get the electronic self energy we apply the
moment-conserving decoupling approximation (MCDA) for suitably defined
Green functions \cite{NRMJ97,schiller01:_kondo}. In the set of equations
which constitute 
the formal solution for the electronic self energy there appear the just
mentioned 4f-spin correlations. Although the MCDA is partly based on
an intuitive ansatz, it has been proved to be quite a reliable approach
to the sophisticated many-body problem of the
multiband-KLM. Interpolating self energy  approaches \cite{nrrmk03}, which
fulfill a maximum number of exact limiting cases, as well as a
systematic projection operator method \cite{HiNo03} yield essentially identical results.
We do not present details of the MCDA but refer the reader to
ref.\ \cite{NRMJ97,schiller01:_kondo}. 

In summary our theory contains only one parameter, namely the
exchange coupling $J$. We assume that it is the same value for the film as
for bulk-EuS, $J=0.23 \mathrm{eV}$ \cite{MuNo02}. Note that $J$ is not really a
free parameter but being read off from the LDA band structure
calculation.

\begin{figure}[htbp]
    \includegraphics[clip=,angle=-90,width=0.95\linewidth]{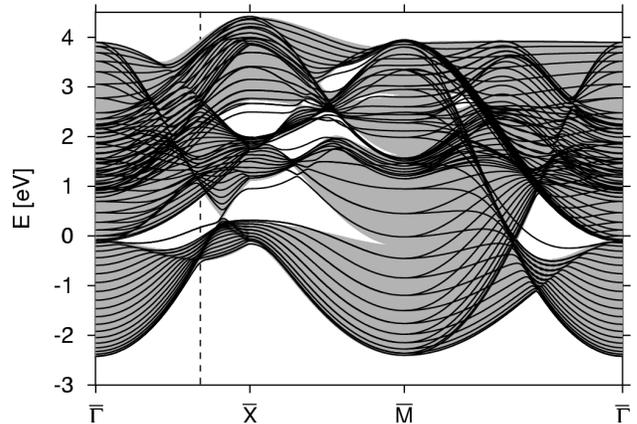}
    \caption{Projected unoccupied (5d)  $\uparrow$ band structure ($T=0\mathrm{K}$) for a
      20 layer 
      EuS(100) film. The shaded region belongs to the respective
      projected band structure of bulk EuS. The vertical broken line identifies
      the position in k-space of the  surface state near the energy
      zero, which is investigated in Fig.~\ref{fig:spectralweight}. The
      energy zero has been chosen arbitrarily.
    \label{fig:EuSbs}}
\end{figure}
\begin{figure}[htbp]
    \includegraphics[height=0.9\linewidth,clip=,angle=-90]{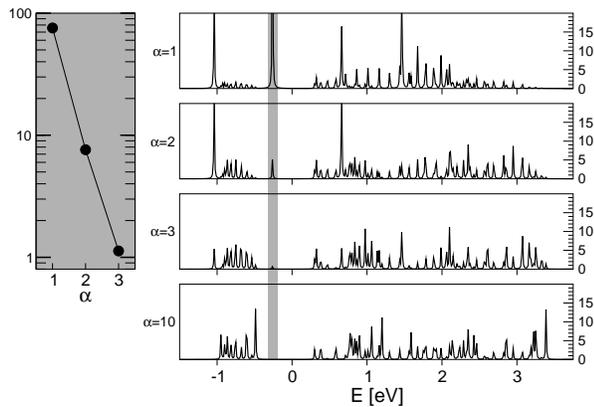}
    \caption{Spectral density ($T=0$, $\sigma=\uparrow$) as function of
    energy for the wave vector at
    $\scriptstyle\frac{2}{3}\textstyle\overline{\Gamma X}$ 
    (dashed line in Fig.~\ref{fig:EuSbs}) and for different layers
    ($\alpha=1,2,3,10$) 
    of a 20-layer EuS(100)
    film. The inset shows on a logarithmic scale the spectral weight of
    the state  within the grey column for the surface and the two layers
    next to the surface. $\alpha = 1 (10)$ means surface (middle) layer.
    \label{fig:spectralweight}}
\end{figure}
\begin{figure}[htbp]
    \includegraphics[width=0.8\linewidth,clip=]{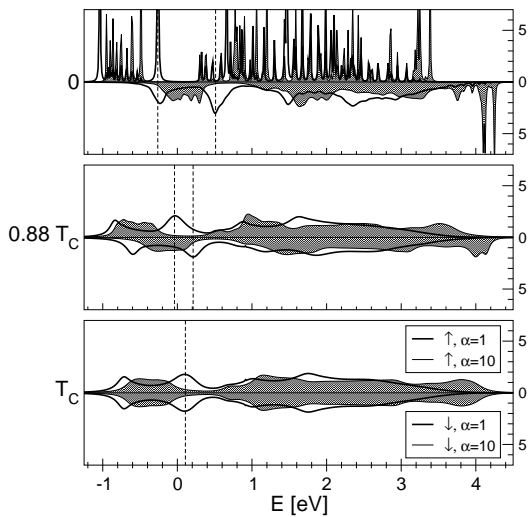}
    \caption{Spectral density of a 20-layer EuS(100) film at
      $\frac{2}{3}\overline{\Gamma X}$ as function of energy and that
      for three different temperatures ($T=0, 0.88T_{\mathrm{C}},
      T_{\mathrm{C}}$). Upper 
      halves for the up-spin spectrum, lower halves for the down-spin
      spectrum. Full lines: surface layer ($\alpha=1$), thin lines
      (shaded regions): middle laver ($\alpha=
      10$). $T_{\mathrm{C}}=16.28 \mathrm{K}$ for 
      the 20-layer film. The vertical broken lines indicate the position
      of the spin-split surface state; left line for $\sigma=\uparrow$,
      right line for $\sigma=\downarrow$.
    \label{fig:Temp}}
\end{figure}

Fig.~\ref{fig:EuSbs} shows the unoccupied (5d) $\uparrow$ band structure of a
20 layer EuS(100) film at $T=0\mathrm{K}$, calculated with the
tight-binding-LMTO 
method. These results represent the single-particle input for our model
calculation. 
Note that the energy zero has been chosen arbitrarily, not at all
coinciding with the Fermi edge. For simplicity we restrict our
presentation to the 5d states only \cite{MuNo02}.
For comparison we also indicate in Fig.~\ref{fig:EuSbs}
the corresponding projected (5d) bulk-band structure, which has the
density of states shown in the upper part of Fig.~2 in \cite{MuNo02}.  
Above all that helps
clearly to identify surface states. A surface state is a state which
appears in the forbidden region where no bulk states occur.
Another characteristic feature of a surface state is the exponential
decay of its 
spectral weight with increasing distance from the surface. To
demonstrate this, we have plotted in Fig.~\ref{fig:spectralweight}
the up-spin spectral density as function of energy for the cut, which is
marked in Fig.~\ref{fig:EuSbs} by a broken line, i.~e.\ for the wave vector at
$\frac{2}{3}\overline{\Gamma X}$. Furthermore, the spectral density is
represented for different layers of the 20-layer film. $\alpha=1$
denotes the surface layer and $\alpha=10$ the middle layer. Note that
the spectral density consists for $T=0$ and $\sigma=\uparrow$
exclusively of
delta-functions representing quasiparticles of infinite lifetimes. Only
to visualize the spectrum we have added a small 
imaginary part to the self energy getting therewith spectral density
peaks of finite widths. 
The structure within the shaded region is obviously a surface state. Its
spectral 
weight decreases exponentially with increasing distance of the
respective layer from the surface. Another surface state appears at the
bottom of the spectrum. We interpret this surface state as the analogue
to that of EuO 
which has been speculated in
ref.~\cite{schiller01:_predic_euo} to give rise to a surface
metal-insulator transition when cooling down from $T_{\mathrm{C}}$ to
$T=0$. 
%
Fig.~\ref{fig:Temp} shows the temperature dependence of the spectral
density, again for the wave vector at  $\frac{2}{3}\overline{\Gamma
  X}$ to include the above-discussed surface state. At $T=0$ (ferromagnetic saturation) the spiky structure of the
$\uparrow$ spectral density refers to quasiparticles of infinite
lifetimes. The $\uparrow$ electron has no chance to exchange its spin
with the parallel aligned localized 4f spins. It therefore propagates
through the lattice without any 
scattering. On the other hand, a $\downarrow$ electron can be scattered by
magnon emission or can form a magnetic polaron what gives rise to a
dramatic lifetime broadening of the 
spectrum even at $T=0$. 
We want to stress that the ($T=0,\sigma=\downarrow$) calculation is exact:
For finite temperature, i.~e. finite magnon
densities, the $\uparrow$ electron, too, can be scattered by magnon
absorption with a concomitant spin flip. Such processes lead to a rather
structureless 5d excitation spectrum, which, nevertheless, carries a
distinct layer dependence. The surface state, identified in
Fig.~\ref{fig:spectralweight}, is spin split at $T=0\mathrm{K}$ by some
$0.8\mathrm{eV}$
due to the exchange coupling to the ferromagnetic 4f moment
system. The induced exchange splitting is strongly temperature-dependent
with a \textit{Stoner-like} collapsing for $T\rightarrow
T_{\mathrm{C}}$. The same 
behavior is shown up by the surface state at the bottom of the
spectrum. An analogous feature has been reported for EuO in
ref.\ \cite{schiller01:_predic_euo}. This distinct temperature dependence
in the ferromagnetic phase appears at first glance somewhat astonishing
because it happens to the unoccupied and \textit{"a priori"}
uncorrelated $5d$ energy states of semiconducting EuS.

By a combination of an LDA-band structure calculation with a many-body
evaluation of the multiband Kondo-lattice model a pronounced temperature
dependence of the (empty) $5d$ conduction bands of EuS(100) films could
be demonstrated. Exchange split magnetic surface states show up a
\textit{Stoner-like} collapsing for $T\rightarrow T_{\mathrm{C}}$. The Curie
temperature exhibits a characteristic film-thickness dependence
understandable as a typical finite size effect.

Financial support by the SFB 290 of the \textit{"Deutsche
Forschungsgemeinschaft"} is gratefully acknowledged. 
\bibliographystyle{apsrev}

\end{document}